\documentclass[runningheads,envcountsame,a4paper,12pt]{llncs}

\pdfoutput=1

\usepackage{epsfig}
\usepackage{subfigure}
\usepackage{calc}
\usepackage{amssymb}
\usepackage{amstext}
\usepackage{amsmath}
\usepackage{color}
\usepackage{graphicx}
\usepackage[colorlinks, citecolor=blue, linkcolor=blue,
urlcolor=blue]{hyperref}
\usepackage{amsfonts}\usepackage{mathrsfs}
\usepackage[normalem]{ulem}
\usepackage{mathptmx}
\usepackage{bbold}
\usepackage[LY1]{fontenc}

\newlength\nesw
\def\neswarrow{\nearrow\hspace{-\nesw}\swarrow}

\newlength\nwse
\def\nwsearrow{\nwarrow\hspace{-\nwse}\searrow}

\newcommand{\ket}[1]{|{#1} \rangle}

\newcommand\bket[1]{|\mathfrak{#1}\rangle}

\begin{document}
\title{From Practice to Theory: The ``Bright Illumination'' Attack
on Quantum Key Distribution Systems\thanks{The work of T.M. and R.L.
was partly supported by the Israeli MOD Research and Technology Unit,
and by the Gerald Schwartz \& Heather Reisman Foundation.}}
\titlerunning{From Practice to Theory: The ``Bright Illumination''
Attack on QKD Systems}
\toctitle{From Practice to Theory: The ``Bright Illumination'' Attack
on Quantum Key Distribution Systems}
\author{Rotem Liss \and Tal Mor}
\institute{Computer Science Department,
Technion---Israel Institute of Technology, Haifa 3200003, Israel \\
\email{$\{$rotemliss,talmo$\}$@cs.technion.ac.il}}
\authorrunning{R. Liss and T. Mor}
\tocauthor{Rotem~Liss and Tal~Mor}

\maketitle

\begin{abstract}
The ``Bright Illumination'' attack~\cite{makarov10} 
is a practical attack, fully implementable against
quantum key distribution systems.
In contrast to almost all developments in quantum information processing
(for example, Shor's factorization algorithm, quantum teleportation,
Bennett-Brassard (BB84) quantum key distribution,
the ``Photon-Number Splitting'' attack, and many other examples),
for which theory has been proposed decades before a proper
implementation, the ``Bright Illumination'' attack preceded
any sign or hint of a theoretical prediction. 

Here we explain
how the ``Reversed-Space'' methodology of attacks,
complementary to the notion of ``quantum side-channel attacks''
(which is analogous to a similar term in
``classical''---namely, non-quantum---computer security), has missed
the opportunity of predicting the ``Bright Illumination'' attack.

\end{abstract}

\section{Introduction}
In the area of quantum information processing,
theory usually precedes experiment.
For example, the BB84 protocol for quantum key distribution (QKD)
was suggested in 1984~\cite{bb84},
five years before it was implemented~\cite{bb84_experiment},
and it still cannot be implemented in a perfectly secure way
even today~\cite{qkd_security_review,qkd_black_paper}.
The ``Photon-Number Splitting'' attack was suggested in
2000~\cite{BLMS00,BLMS00_eurocrypt}, but it is not implementable today.
Quantum computing was suggested
in the 1980s (see, e.g.,~\cite{deutsch85}),
but no useful and universal quantum computer
(with a large number of clean qubits) has been implemented until
today~\cite{quantum_computers_NISQ18}.
The same applies to Shor's factorization
algorithm~\cite{shor_algo94,shor_algo99},
to quantum teleportation~\cite{teleport}
(at least to some extent; see also~\cite{nv_teleport14}),
and to many other examples.

In contrast to the above examples, the ``Bright Illumination'' attack
against practical QKD systems was presented and fully implemented
in 2010~\cite{makarov10}, \emph{prior} to any theoretical prediction
of the possibility of such an attack.
Here we ask the question: could the ``Bright Illumination''
attack have been theoretically predicted?

Quantum key distribution (QKD) makes it possible for two parties,
Alice and Bob, to agree on a shared secret key. This task, that
is impossible for two parties using only classical communication,
is made possible by taking advantage of quantum phenomena:
Alice and Bob use an insecure quantum channel
and an authenticated (unjammable) classical channel.
The resulting key is secure even against an adversary (Eve) that
may use the most general attacks allowed by quantum physics,
and remains secret indefinitely, even if Eve has unlimited computing 
power.

For example, in the BB84 QKD protocol~\cite{bb84}, Alice sends to Bob
$N$ qubits, each of them randomly chosen from the set of quantum states
$\{\bket{0}, \bket{1},
\bket{+} \triangleq \frac{\bket{0} + \bket{1}}{\sqrt{2}},
\bket{-} \triangleq \frac{\bket{0} - \bket{1}}{\sqrt{2}}\}$,
and Bob measures each of them either in the computational basis
$\{\bket{0}, \bket{1}\}$ or in the Hadamard basis
$\{\bket{+}, \bket{-}\}$, chosen randomly.
Thereafter, Alice and Bob post-process the results
by using the classical channel.
If Alice and Bob use matching bases, they share a bit
(unless there is some noise or eavesdropping);
if they use mismatching bases, Bob's results are random.
Alice and Bob reveal their basis choices, and discard
the bits for which they used mismatching bases. After that, they
publicly reveal a random subset of their bits in order to estimate
the error rate (then discarding those exposed test bits),
aborting the protocol if the error rate is too high;
and they perform error correction and privacy amplification processes,
obtaining a completely secret, identical final key.

The security promises of QKD are true in theory,
but its \emph{practical} security is far from being guaranteed.
The practical implementations of QKD use realistic photons;
therefore, they deviate from the theoretical protocols,
which use ideal qubits. These deviations make possible
various attacks~\cite{qkd_security_review,qkd_security_practical},
related to the idea of ``side-channel attacks''
in classical (i.e., non-quantum) computer security.

For example, the ``Photon-Number Splitting'' (PNS) 
attack~\cite{BLMS00,BLMS00_eurocrypt}
(see Subsection~\ref{subsec:fock}) takes advantage of
some specific imperfections: while the quantum state sent
by Alice \emph{should} be encoded in a single photon,
Eve exploits the fact that in most implementations, Alice
sometimes sends to Bob more than one photon (e.g., two photons).
The PNS attack was found using
more realistic notations---the Fock space notations;
the main insight of~\cite{BLMS00,BLMS00_eurocrypt} 
is that using proper notations is vital,
both when theoretically searching for possible loopholes
and attacks against QKD,
and when attempting to prove its security.

The ``Bright Illumination'' practical attack~\cite{makarov10} uses a
weakness of Bob's measurement devices that allows Eve to ``blind'' them
and fully control Bob's measurement results. Eve can then
get full information on the secret key, without inducing any error.

In Section~\ref{exp_imper} we explain
experimental QKD systems and their weaknesses:
we introduce the Fock space notations,
the ``Photon-Number Splitting'' (PNS) attack,
and two imperfections of Bob's detection process.
In Sections~\ref{bia} and~\ref{rsa} we describe the
practical ``Bright Illumination'' attack and
the ``Reversed-Space'' methodology of
attacks~\cite{reversed12,fixed14,reversed16}, respectively,
and in Section~\ref{sec:p2t} we bring together all the above notions
for explaining the theory underlying
the ``Bright Illumination'' attack.
As an important side issue,
in Section~\ref{side-channel} we describe the notion of
``quantum side-channel attacks'', partially related to all the above.
We conclude that while the ``Bright Illumination''
attack is not a ``side-channel'' attack, it
can be modeled as a ``Reversed-Space'' attack~\cite{reversed12}:
this attack and similar attacks
could and should have been proposed or anticipated by theoreticians.

\section{\label{exp_imper}Experimental QKD, Imperfections,
and the Fock Space Notations}
The BB84 protocol may be experimentally implemented in a
``polarization-based'' implementation, that we can model as follows:
Alice's quantum states, that are sent to Bob, are single photons
whose polarizations encode the quantum states.
The four possible states to be sent by Alice are $\bket{0}$, $\bket{1}$,
$\bket{+}$, and $\bket{-}$,
where $\bket{0} = \ket{\leftrightarrow}$
(a single photon in the horizontal polarization)
and $\bket{1} = \ket{\updownarrow}$
(a single photon in the vertical polarization).
The states $\bket{+} = \ket{\neswarrow}$ and
$\bket{-} = \ket{\nwsearrow}$ correspond to
orthogonal diagonal polarizations.

For measuring the incoming photons,
Bob uses a polarizing beam splitter (PBS) and two detectors.
Bob actively configures the PBS for choosing
his random measurement basis (the computational basis
$\{\bket{0}, \bket{1}\}$ or the Hadamard basis
$\{\bket{+}, \bket{-}\}$).
If the PBS is configured for measurement in the computational basis,
it sends any \emph{horizontally} polarized photon to one arm
and sends any \emph{vertically} polarized photon to the other arm.
In the end of each arm, a detector is placed, which clicks whenever
it detects a photon.
Therefore, the detector in the first arm clicks \emph{only}
if the $\bket{0}$ qubit state is detected,
and the detector in the second arm clicks \emph{only}
if the $\bket{1}$ qubit state is detected.
A \emph{diagonally} polarized photon
(i.e., $\bket{+} = \ket{\neswarrow}$ or $\bket{-} = \ket{\nwsearrow}$)
would cause exactly one of the detectors (uniformly random) to click.
Similarly, if the PBS is configured
for measurement in the Hadamard basis, it distinguishes 
$\bket{+}$ from $\bket{-}$. This implementation may be slow, because
Bob needs to randomly choose a basis for each arriving photon.

A more practical---yet imperfect---variant
of this implementation uses a ``passive'' basis choice
(e.g.,~\cite{fixed_example02}).
This variant uses one polarization-independent beam splitter,
two PBSs, and \emph{four} detectors.
In this variant, the polarization-independent beam splitter
randomly sends each photon to one arm
or to another. A photon going to the first arm is then measured
(as described above) in the computational basis,
while a photon going to the second arm is measured (as
described above) in the Hadamard basis.
This ``passive'' variant is exposed to various attacks; see
Section~\ref{rsa}.

\subsection{\label{subsec:fock}The Fock Space Notations
and the ``Photon-Number Splitting'' (PNS) Attack}
We use the Fock space notations for describing practical QKD systems:
\begin{itemize}
\item In the simplest case, there are $k \ge 0$ photons,
and all these photons belong to \emph{one} photonic mode.
The Fock state $\ket{k}$ represents $k$ photons in this single mode:
for example, $\ket{0}$ is the vacuum state,
representing no photons in that mode;
$\ket{1}$ represents one photon in that mode;
$\ket{2}$ represents two photons in that mode; and so on.
\item For describing several different \emph{pulses} of photons
(for example, photons traveling on different arms or
at different time bins,
or any other \emph{external} degree of freedom),
we need several photonic \emph{modes}.
For example, if we assume a single photon in two pulses
(and, thus, in two modes), we can describe 
a qubit\footnote{The notations $\bket{0}, \bket{1}, \bket{\pm}$
are used for the standard qubit (in a two-dimensional Hilbert space).}:
for the computational basis $\{\bket{0}, \bket{1}\}$ of a single qubit,
we write $\bket{0} = \ket{0} \otimes \ket{1} \equiv \ket{0} \ket{1}$
and $\bket{1} = \ket{1} \otimes \ket{0} \equiv \ket{1} \ket{0}$.
(Those two modes are mathematically described
using a tensor product, but we omit the $\otimes$ sign for brevity.)
A superposition, too, describes a single photon in those two pulses:
for example, the Hadamard basis states are
$\bket{\pm} = \frac{\ket{0}\ket{1} \pm \ket{1}\ket{0}}{\sqrt{2}}$.
\item More generally, if we have $k = k_1 + k_0$ photons
in two different pulses (two modes),
where $k_1$ photons are in one pulse
and $k_0$ photons are in the other pulse,
we write $\ket{k_1} \ket{k_0}$.
Subscripts are added for specifying the types of pulses---for example,
$\ket{k_1}_{t_1} \ket{k_0}_{t_0}$ for the two time bins
$t_1, t_0$, or $\ket{k_1}_A \ket{k_0}_B$ for the two arms $A,B$.
\item For describing more than two pulses
(namely, more than two modes), we use generalized notations:
for example, $k = k_2 + k_1 + k_0$ photons in three time bins
are denoted $\ket{k_2}_{t_2} \ket{k_1}_{t_1} \ket{k_0}_{t_0}$.
In particular, the vacuum state (absence of photons) is denoted
$\ket{0}$ for one mode, $\ket{0} \ket{0}$ for two modes,
$\ket{0} \ket{0} \ket{0}$ for three modes, and so on.
\end{itemize}

The above notations assume the photon \emph{polarizations}
(which are an \emph{internal} degree of freedom) to be identical
for all $k$ photons. However, a single photon in a single
pulse generally has two orthogonal polarizations:
horizontal $\leftrightarrow$ and vertical $\updownarrow$.
For each pulse, the two polarizations are described as two modes;
therefore, $m$ pulses mean $2m$ modes.

In this paper, we denote \emph{polarization} modes of $k=k_1+k_0$
photons by $\ket{k_1,k_0}$ (without any subscript),
and denote only \emph{pulse} modes by $\ket{k_1}\ket{k_0}$
(always with subscripts). Thus:
\begin{itemize}
\item For a \emph{single} pulse, the two \emph{polarization} modes
describe a qubit if there is exactly \emph{one photon} in the pulse.
The computational basis states are $\bket{0} = \ket{0,1}$
(representing \emph{one} photon in the horizontal polarization mode and
\emph{zero} photons in the vertical polarization mode) and
$\bket{1} = \ket{1,0}$
(where the single photon is in the vertical mode).
\item Similarly to the above,
we can also describe: (a)~superpositions;
(b)~the state $\ket{k_1,k_0}$ of $k=k_1+k_0$ photons
in those two polarization modes ($k_1$ photons in the vertical mode
and $k_0$ photons in the horizontal mode);
and (c)~the vacuum state $\ket{0,0}$.
\end{itemize}

We have seen that the Fock space notations extend \emph{much} beyond
the ideal single-qubit world,
which is represented by the two-dimensional
space $\text{Span}\{\bket{0}, \bket{1}\}$.
Ideally, in BB84, Alice should send a qubit
in this two-dimensional space;
however, in practice, Alice sometimes sends states
in a higher-dimensional Fock space.

The ``Photon-Number Splitting'' (PNS)
attack~\cite{BLMS00,BLMS00_eurocrypt} (which showed all QKD experiments
done until around 2000 to be insecure) is based on
analyzing the \emph{six}-dimensional Hilbert space
$\text{Span}\{\ket{0,0}, \ket{0,1}, \ket{1,0}, \ket{0,2}, \ket{2,0},
\ket{1,1}\}$, which represents all typical pulses with two
polarizations if we can neglect the case of
three or more photons---namely, if we assume $k_1+k_0 \le 2$.
The PNS attack is based
on three observations~\cite{BLMS00,BLMS00_eurocrypt}:
(a)~Alice sometimes sends \emph{two-photon pulses}
in one of the four allowed polarizations;
(b)~Eve can, in principle, distinguish a two-photon pulse
from a single-photon pulse without influencing the polarizations; and
(c)~Eve can, in principle, split such a two-photon pulse
into two pulses,
each containing a single photon, without influencing the polarizations.
Thus, Eve can ``steal'' a single photon from each such two-photon pulse
(without influencing the other photon), save it,
and, after learning the basis,
get full information about this pulse without being noticed.
This attack could have been detrimental to the security of QKD, but
counter-measures~\cite{decoy_hwang03,decoy_wang05,decoy_LMC05,SARG04}
have been found later.

\subsection{\label{subsec:imperfect}Imperfections of Bob's Detectors}
Two important examples of imperfections (see~\cite{reversed16})
are highly relevant to various ``Reversed-Space'' attacks.
As we show in this paper, those two imperfections must be
\emph{combined} for understanding the ``Bright Illumination'' attack.

\paragraph*{Imperfection 1:}
Our realistic assumption, which is true for standard detectors in QKD
implementations, is that Bob's detectors cannot \emph{count}
the number of photons in a pulse.
Thus, they cannot distinguish \emph{all} Fock states
$\ket{k}$ from one another, but can only distinguish
the Fock state $\ket{0}$ (a lack of photons) from
the Fock states $\{\ket{k} : k \ge 1\}$.
Namely, standard detectors can only decide
whether the mode is empty ($k = 0$)
or has at least one photon ($k > 0$). In contrast, we assume that Eve
can (in principle) do anything allowed by the laws of quantum physics;
in particular, Eve may have such ``photon counters''.

In particular, let us assume that there are \emph{two} pulses,
each of them consisting of a single mode.
Bob cannot know whether a pulse contains one photon or
two photons; therefore, he cannot distinguish between
$\ket{1}\ket{0}$ and $\ket{2}\ket{0}$ (and, similarly, he cannot 
distinguish between $\ket{0}\ket{1}$ and $\ket{0}\ket{2}$).
For example, assume that Alice sends the $\ket{1}\ket{0}$ state
(a qubit) to Bob, and Eve replaces Alice's state by $\ket{2}\ket{0}$
and sends it to Bob instead (or, similarly, assume that Eve replaces
$\ket{0}\ket{1}$ by $\ket{0}\ket{2}$).
In this case, Bob cannot notice the change, and 
no error can occur; still, Bob got a state he had not expected to get.
It may be possible for Eve to take advantage of this fact in
a fully-designed attack.

\paragraph*{Imperfection 2:}
Our realistic assumption is that Bob cannot know exactly \emph{when}
the photon he measured arrived. For example (in a polarization-based
implementation):
\begin{itemize}
\item Alice's ideal qubit arrives at time $t$
(states denoted $\ket{0,1}_t \ket{0,0}_{t+\delta} \;$,
$\ket{1,0}_t \ket{0,0}_{t+\delta}$).
\item Eve's photon may arrive at time $t+\delta$
(states denoted $\ket{0,0}_t \ket{0,1}_{t+\delta} \;$,
$\ket{0,0}_t \ket{1,0}_{t+\delta}$).
\end{itemize}
Again, Eve may take advantage of this fact in a fully-designed attack.

Similar imperfections can be found if Bob cannot know exactly
what the \emph{wavelength} of the photon is,
or \emph{where} the photon arrives.

\paragraph*{The conceptual difference between the two imperfections}
is in whether Bob can (ideally) avoid measuring the extra states
sent by Eve, or not:
\begin{itemize}
\item In Imperfection~1, Eve may send more than one photon, and Bob 
must measure the state
(while he cannot count the number of photons using current technology).
\item In Imperfection~2, Eve sends states in two separate subsystems.
Bob can, in principle, ignore the ``wrong'' subsystem
in case he knows for sure it has not been sent by Alice.
\end{itemize}

\section{\label{bia}The ``Bright Illumination'' Attack}
The ``Bright Illumination'' blinding attack~\cite{makarov10}
works against QKD systems that use Avalanche Photodiodes (APDs)
as Bob's detectors. As an example, we describe below the implementation
of this attack against a system implementing the BB84 protocol in
a polarization-based scheme, but it is important to note 
that the attack can be 
adapted to most QKD protocols and implementations
that use APDs~\cite{makarov10}.

The APDs can be operated in two ``modes of operation'':
the ``linear mode'' that detects
only a light beam above a specific power threshold,
and ``Geiger mode'' that detects even a single photon
(but cannot count the number of photons).
In this attack, the adversary Eve sends a continuous strong light beam
towards Bob's detectors, causing them to operate
\emph{only} in the linear mode (thus ``blinding'' the detectors).

After Bob's detectors have been blinded (and in parallel to sending
the continuous strong beam, making sure they are kept blind),
Eve performs a ``measure-resend'' attack: she detects the qubit
(single photon)  sent by Alice, measures it in one of the two bases
(exactly as Bob would do),
and sends to Bob a \emph{strong} light beam
depending on the state she measured,
a little above the power threshold of the detectors.
For example, if Eve measures the state $\ket{1,0}$,
she sends to Bob the state $\ket{k,0}$ for $k \gg 1$.
Now, if Bob chooses the same basis as Eve,
he will measure the same result as Eve;
and if Bob chooses a different basis,
he will measure nothing,
because the strong light beam will get split between the two detectors.
This means that Bob will always either measure the same result as Eve
or lose the bit.

In the end, Bob and Eve have exactly the same information, so Eve can
copy Bob's classical post-processing and get the same final key as
Alice and Bob do. Moreover, Eve's attack causes no detectable
disturbance, because Bob does not know
that his detectors have operated in the wrong mode of operation;
the only effect is a loss rate of 50\% (that is not problematic:
the loss rate for the single photons sent by Alice
is usually much higher, so Eve can cause Bob to get the same loss rate
he expects to get).

This attack was both developed and experimentally demonstrated
against commercial QKD systems by~\cite{makarov10}.
See~\cite{makarov10} for more details and for diagrams.

\section{\label{rsa}``Reversed-Space'' Attacks}
The ``Reversed-Space'' methodology, described
in~\cite{reversed_thesis,reversed16,reversed12},
is a theoretical framework of attacks
exploiting the imperfections of Bob.
This methodology is a special case (easier to analyze)
of the more general methodology of
``Quantum Space'' attacks~\cite{qsa07,reversed_thesis},
that exploits the imperfections of \emph{both} Alice and Bob;
the ``Reversed-Space'' methodology assumes Alice to be ideal
and only exploits Bob's
imperfections~\cite{reversed_thesis,reversed12,fixed14,reversed16}.
(Another special case of a ``Quantum Space'' attack is
the PNS attack~\cite{BLMS00,BLMS00_eurocrypt}
described above.)

In the ideal QKD protocol, Bob expects to get from Alice a state in
the Hilbert space $\mathcal{H}^\mathrm{A}$;
however, in the ``Reversed-Space'' attack, Bob gets from Eve
an unexpected state, residing in a larger Hilbert space
called the ``space of the protocol'' and denoted by
$\mathcal{H}^\mathrm{P}$.
In principle, Eve could have used a huge space 
$\mathcal{H}'$ such that 
$\mathcal{H}^\mathrm{A} \subseteq
\mathcal{H}^\mathrm{P} \subseteq \mathcal{H}'$:
the huge Hilbert space $\mathcal{H}'$ consists of \emph{all} the quantum
states that Eve \emph{can possibly} send to Bob, but it is too large,
and most of it is irrelevant.

Because ``Reversed-Space'' attacks assume
a ``perfect Alice'' (sending prefect qubits), it is usually easy 
to find the \emph{relevant} subspace
$\mathcal{H}^\mathrm{P}$, as we demonstrate
by three examples below; $\mathcal{H}^\mathrm{P}$
is only enlarged (relative to the ideal space $\mathcal{H}^\mathrm{A}$)
by Bob's imperfections. Therefore, $\mathcal{H}^\mathrm{P}$ is
the space that includes all the states that may be useful for Eve
to send to Bob.
The space $\mathcal{H}^\mathrm{P}$ is defined by taking
all the possible measurement results of Bob and reversing them in time;
more precisely, it is the span of all the states in
$\mathcal{H}^\mathrm{A}$ \emph{and} all the states that Eve can send
to Bob so that he gets the measurement results she desires.

Whether Bob is aware of it or not, 
his experimental setting treats not only the states in 
$\mathcal{H}^\mathrm{A}$, but all the possible inputs
in the ``space of the protocol'' $\mathcal{H}^\mathrm{P}$.
Bob then classifies them into three classes:
(1)~valid states from Alice, (2)~losses, and (3)~invalid states.
\emph{Valid states} are always treated in conventional security
analysis: a random subset is compared with Alice for estimating
the error rate, and then the final key is obtained using
the error correction and privacy amplification processes.
\emph{Losses} are expected, and they are not counted as noise.
\emph{Invalid states} are usually counted as errors (noise),
but they do not appear in ideal analyses of ideal protocols.
We note that loss rate and error rate are computed separately:
the error rate must be small (e.g., around 10\%) for the protocol
not to be aborted by Alice and Bob, while the loss rate can be
much higher (even higher than 99\%). Any ``Reversed-Space'' attack takes
advantage of the possibility that Bob treats some states in
$\mathcal{H}^\mathrm{P}$ in the wrong way,
because he does not expect to get those states.

Eve's attack is called ``Reversed-Space'' because Eve can devise
her attack by looking at Bob's possible measurement results:
Eve finds a measurement result she wants to be obtained by Bob
(because he interprets it in a way desired by her)
and reverses the measurement result in time for finding
the state in $\mathcal{H}^\mathrm{P}$ she should send to Bob.
In particular, if Bob applies the unitary operation
$\mathcal{U}_\mathrm{B}$ on his state prior to his measurement,
Eve should apply the inverted operation
$\mathcal{U}_\mathrm{B}^{-1} = \mathcal{U}_\mathrm{B}^\dagger$ to each 
state corresponding to each possible
measurement outcome of Bob.

We present three examples of ``Reversed-Space'' attacks.
For simplicity, we only consider BB84
implemented in a polarization-based scheme (as described in
Section~\ref{exp_imper}), but the attacks may be generalized
to other implementations, too.
We emphasize that all three examples have been chosen to
satisfy two conditions, also satisfied by
the ``Bright Illumination'' attack:
(a)~Eve performs a ``measure-resend'' attack in a basis she chooses
randomly, and (b)~it is possible for Eve to get full information
without inducing noise.

\paragraph{Example 1 (a special case of the ``Trojan Pony'' 
attack~\cite{gllp04}):}
This example exploits Imperfection~1 and assumes Bob uses
an ``active'' basis choice (see Section~\ref{exp_imper} for both).

In this attack, Eve performs a ``measure-resend'' attack---namely,
she measures each qubit state sent from Alice to Bob in a random basis,
and resends ``it'' towards Bob. However, instead of resending it
as a single photon,
she resends a huge number of photons towards Bob:
she sends many \emph{identical} photons,
all with the same polarization as the state she measured
($\bket{0}$, $\bket{1}$, $\bket{+}$, or $\bket{-}$).
If Bob chooses the same basis as Eve,
he will get the same result as her,
because Imperfection~1 causes his system to treat
the incoming states $\ket{0,k}$ and $\ket{k,0}$ (for any $k \ge 1$)
as if they were $\ket{0,1}$ and $\ket{1,0}$, respectively;
but if he chooses a different basis from Eve,
both of his detectors will (almost surely) click.
If Bob decides to treat
this \emph{invalid} event (a two-detector click) as an ``error'',
the error rate will be around 50\%,
so Alice and Bob will abort the protocol;
but if he naively decides to treat this event as a ``loss'',
Eve can get full information without inducing errors.

Alice sends an ideal qubit (a single photon),
while Eve may send any number of photons. Therefore,
using the Fock space notations,
$\mathcal{H}^\mathrm{A} = \mathcal{H}_\mathrm{2} \triangleq
\text{Span}\{\ket{0,1}, \ket{1,0}\}$ and $\mathcal{H}^\mathrm{P} =
\text{Span}\{\ket{m_1,m_0} : m_1, m_0 \ge 0\}$.

\paragraph{Example 2 (a special case of the ``Faked States'' 
attack~\cite{faked_states05,faked_states06,reversed_thesis}):}
This attack exploits Imperfection~2 (Section~\ref{exp_imper}).
We assume that Bob has four detectors
(namely, that he uses the ``passive'' basis choice
variant of the polarization-based encoding;
see Section~\ref{exp_imper}),
and that his detectors have different (but overlapping)
\emph{time gates} during which they are sensitive:
given the three different times $t_0 < t_{1/2} < t_1$,
the detectors for the computational basis are sensitive only to pulses
sent at $t_0$ or $t_{1/2}$ (or in between),
and the detectors for the Hadamard basis are sensitive only to pulses
sent at $t_{1/2}$ or $t_1$ (or in between).
Alice normally sends her pulses at $t_{1/2}$
(when both detectors are sensitive), but Eve may send her
pulses at $t_0$, $t_{1/2}$, or $t_1$.

Eve performs a ``measure-resend'' attack by measuring Alice's state
in a random basis, and resending it towards Bob as follows:
if Eve measures in the computational basis,
she resends the state at time $t_0$; and if she measures in
the Hadamard basis, she resends the state at time $t_1$.
Therefore, Bob gets the same result as Eve if he measures in the
same basis as hers,
but he gets a loss otherwise (because Bob's detectors
for the other basis are not sensitive at that timing).
This means that Eve gets full information without inducing any error.

Using the same notations as in Imperfection~2,
the state $\ket{m_1,m_0}_{t_0} \ket{n_1,n_0}_{t_{1/2}} 
\ket{o_1,o_0}_{t_1}$ consists of the Fock states
$\ket{m_1,m_0}$ sent at time $t_0$, $\ket{n_1,n_0}$
sent at time $t_{1/2}$, and $\ket{o_1,o_0}$ sent at time
$t_1$. Alice sends an ideal qubit (a single photon at time $t_{1/2}$),
while Eve may send a single photon at any of the times $t_0$,
$t_{1/2}$, or $t_1$, or a superposition. 

Therefore,
$\mathcal{H}^\mathrm{A} = \mathcal{H}_\mathrm{2} \triangleq 
\text{Span}\{\ket{0,0}_{t_0} \ket{0,1}_{t_{1/2}} 
\ket{0,0}_{t_1}~,~
\ket{0,0}_{t_0} \ket{1,0}_{t_{1/2}} 
\ket{0,0}_{t_1}\}$
and
$\mathcal{H}^\mathrm{P} = 
\text{Span}\{\ket{0,1}_{t_0}  \ket{0,0}_{t_{1/2}} 
\ket{0,0}_{t_1}~,~
\ket{1,0}_{t_0}  \ket{0,0}_{t_{1/2}}  \ket{0,0}_{t_1}~,~
\ket{0,0}_{t_0}  \ket{0,1}_{t_{1/2}}  \ket{0,0}_{t_1}~,~\linebreak[4]
\ket{0,0}_{t_0}   \ket{1,0}_{t_{1/2}}   \ket{0,0}_{t_1}~,~
\ket{0,0}_{t_0}  \ket{0,0}_{t_{1/2}}  \ket{0,1}_{t_1}~,~
\ket{0,0}_{t_0}  \ket{0,0}_{t_{1/2}}  \ket{1,0}_{t_1}\}$.

\paragraph{Example 3 (the 
``Fixed Apparatus'' attack~\cite{fixed14})} can be
applied by Eve if Bob uses a ``passive'' basis choice
(Section~\ref{exp_imper}).
In this attack, Eve sends to Bob an unexpected state,
and this state ``forces'' Bob to obtain the basis Eve wants.
This attack makes it possible for Eve
to force Bob choose the same basis as her
(and, therefore, get the same outcome as her),
thus stealing the whole key without inducing any errors or losses.
The attack is only possible if Eve has a one-time access
to Bob's laboratory, because it requires Eve to first compromise
Bob's device (otherwise, she cannot send him that unexpected state).

Assume that Bob uses a polarization-independent beam splitter 
that splits the incoming beam into two different output
arms (as described in Section~\ref{exp_imper}).
This beam splitter has two input arms: a \emph{regular arm},
through which the standard incoming beam comes, and a \emph{blocked
arm}, where the incoming state is always assumed to be
the zero-photon beam $\ket{0,0}$ (the vacuum state of two polarizations).
If Eve can drill a small hole in Bob's device, exactly where the
blocked arm gets its input from, then she can send a beam to the
blocked arm and not only to the standard arm.
It is proved~\cite{fixed14} that Eve can then cause
the beam splitter to choose an output arm to her desire,
instead of choosing a ``random'' arm.
The state $\ket{m_1,m_0}_r 
\ket{n_1,n_0}_b$ consists of the Fock state $\ket{m_1,m_0}$
sent through the \emph{regular arm} of the beam splitter and
the Fock state $\ket{n_1,n_0}$ sent through the \emph{blocked arm}.
Alice sends an ideal qubit (a single photon through the regular arm),
while Eve may send a single photon through any of the two arms
or a superposition. Therefore,
$\mathcal{H}^\mathrm{A} = \mathcal{H}_\mathrm{2} \triangleq 
\text{Span}\{\ket{0,1}_r  \ket{0,0}_b~,~
\ket{1,0}_r  \ket{0,0}_b\}$ and
$\mathcal{H}^\mathrm{P} = 
\text{Span}\{\ket{0,1}_r  \ket{0,0}_b~,~
\ket{1,0}_r  \ket{0,0}_b~,~
\ket{0,0}_r  \ket{0,1}_b~,~
\ket{0,0}_r  \ket{1,0}_b\}$.

\section{\label{side-channel} Quantum Side-Channel Attacks}
\paragraph{Shamir's ``Quantum Side-Channel Attack''
on Polarization-Based QKD:}
The following attack was proposed by Adi Shamir in a meeting with one
of the authors (T.M.) around 1996--1999~\cite{shamir90s},
and it may have never been published (but see similar attacks below).
Shamir's attack only applies to QKD implementations that use
``\emph{active}''
basis choice (as opposed to the ``passive'' basis choice, which leads
to the ``Fixed Apparatus'' attack described in Example~3 of
Section~\ref{rsa}).
The attack is related to Imperfection~2 described in
Section~\ref{exp_imper}: Bob's apparatus must be fully or partially
ready to receive Alice's photon before it arrives.
For example, if the photon is supposed to arrive at time $t$,
then Bob's setup is already partially ready at time $t - \delta$;
in particular, Bob decides the \emph{basis choice} and configures
the polarizing beam splitter accordingly before time $t - \delta$.
The attack also assumes that the detectors themselves are still inactive
(blocked) at time $t - \delta$, and are activated just before time $t$.
Therefore, at time $t - \delta$, the polarizing beam splitter is
already configured to match the required basis (the computational basis
or the Hadamard basis), while the detectors are still blocked.

Eve's attack is sending a strong pulse at time $t - \delta$,
that hits the polarizing beam splitter (but not the blocked detectors)
and gets reflected back to Eve, containing full or partial information
on the direction of the polarizing beam splitter---and, thus,
on the basis choice.
Assuming Eve gets the information on Bob's basis choice
\emph{before} she receives Alice's pulse, Eve could employ
the following full attack: Eve measures the
photon coming from Alice \emph{in the same basis chosen by Bob},
learns the qubit's value,
and resends to Bob the resulting state (in the same basis),
obtaining full information while inducing no errors and no losses.

One can suggest two ways to possibly prevent the attack:
(a)~opening the detection window (activating the detectors)
\emph{shortly} after the polarizing beam splitter is configured
according to the basis choice (if the time difference is
sufficiently short, Eve cannot find Bob's basis choice on
time for employing the full attack); or (b)~blocking access to
the polarizing beam splitter until the detectors are activated
(although this solution may be hard to implement).

As we explain in Section~\ref{sec:p2t}, the ``Bright Illumination''
attack could have been predicted by adding Imperfection~1 described in
Section~\ref{exp_imper} (namely, detection of multi-photon pulses)
to the above idea of a strong pulse sent at time $t - \delta$
towards Bob (i.e., Imperfection~2, as already discussed here)
and using the Fock space notations.

\paragraph{``Conventional Optical Eavesdropping'' and
``Quantum Side-Channel Attacks'':}
Other attacks, similar to Shamir's attack,
have been independently developed---for example,
the ``Large Pulse'' attack~\cite{large_pulse01},
which attacks both Alice's and Bob's set-ups.
As written in~\cite{large_pulse01}:
``This [large pulse] attack is one of the possible methods of
conventional optical eavesdropping, a new strategy of eavesdropping
on quantum cryptosystems, which eliminates the need of
immediate interaction with transmitted quantum states.
It allows the eavesdropper to avoid inducing transmission errors
that disclose her presence to the legal users.''

Instead of restricting ourselves to ``conventional
optical eavesdropping on quantum cryptosystems'',
we make use of a different sentence
from~\cite{large_pulse01}---``eavesdropping on quantum cryptosystems
which eliminates the need of immediate interaction with transmitted
quantum states''---and we define ``quantum side-channel attacks''
as follows:
\begin{description}
\item A \emph{quantum side-channel attack} is any 
eavesdropping strategy which eliminates the need of any
immediate interaction with the transmitted quantum states.
\end{description}

According to the above definition, both Shamir's attack and
the ``Large Pulse'' attack are ``quantum side-channel attacks'',
because they attack the devices and not the quantum states themselves.
On the other hand, the ``Reversed-Space'' attacks and
the ``Quantum Space'' attacks (see Section~\ref{rsa})
can be fully described using a proper description of
the QKD protocol, which uses the Fock space notations;
therefore, they should \emph{not} be considered
as ``quantum side-channel attacks''.
In fact, we can say they are \emph{complementary}
to ``quantum side-channel attacks'',
and we name them ``\emph{state}-channel attacks''.

In a classical communication world, the notion of ``side-channel
attacks'' makes use of any information leaked by the \emph{physical}
execution of the algorithm (see, for example,~\cite{side_channel07}).
Accordingly, other researchers (e.g.,~\cite{qkd_security_practical})
have chosen to adopt a wide definition of ``quantum side-channels'',
which also includes the ``Photon-Number Splitting'' attack and
many other practical attacks.
However, we prefer to take a narrower view of ``quantum side-channel
attacks'', as explained above.

\section{\label{sec:p2t}From Practice to Theory: The Possibility of
Predicting the ``Bright Illumination'' Attack}
The ``Bright Illumination'' attack could have been predicted,
because it simply combines Imperfections~1 and~2 that were described
in Section~\ref{exp_imper}: namely, detecting many photons at time
$t - \delta$, while the single ``information'' photon
should have arrived at time $t$.
In some sense, it seems to merge a ``Reversed-Space'' attack and
a ``quantum side-channel attack'', because it attacks both
the transmitted quantum states and the detectors themselves.
However, because Bob's detectors are fully exposed
to Eve at both times $t$ and $t-\delta$
(unlike the ``Large Pulse'' attack~\cite{large_pulse01},
where the detectors are not exposed at time $t - \delta$),
we see the ``Bright Illumination'' attack
as a special (and fascinating) case of ``Reversed-Space'' attack,
and not as a ``quantum side-channel attack''.

The ``Bright Illumination'' attack is made possible by
a \emph{lack of information} on the 
``space of the protocol'' $\mathcal{H}^\mathrm{P}$:
Eve sends many photons (as in Imperfection~1) at time $t - \delta$
(as in Imperfection~2), and Bob does not notice her disruption
because he cannot \emph{count} the number of photons and
cannot \emph{block} the detectors at time $t - \delta$.

For preventing all the possible attacks and proving full security,
it must be known how Bob's detectors
treat \emph{any} number $k$ of photons sent to him by Eve,
and it must also be known how Bob's detectors treat multiple pulses.
In particular, a detector definitely cannot operate properly
in the hypothetical scenario where an infinite 
number of photons (with infinite energy) arrives as its input.
A potentially secure system must have an estimated threshold $N$,
such that if $k \lesssim N$ photons arrive,
they are correctly measured by the detectors
(treated as one photon), and if $k \gtrsim N$ photons arrive,
the measurement result is clearly invalid and is known to Bob
(for example, smoke comes out of the detectors,
or the detectors are burned). $N$ is estimated,
so there is a small unknown range near it.

Prior to the ``Bright Illumination'' attack, it seems that
no systematic effort has been invested in finding or approximating
the threshold $N$ and characterizing the detectors' behavior
on \emph{all} possible inputs (any number of photons $k$).
A proper ``Reversed-Space'' analysis would have suggested that
experimentalists \emph{must} check what $N$ is
and fully analyze the behavior of Bob's detectors
on each quantum state;
such an analysis would then have found the ``space of the protocol''
$\mathcal{H}^\mathrm{P}$ which is available for Eve's attack.

A careful ``Reversed-Space'' analysis---if it had been
carried out---would then have found that instead of \emph{one}
estimated threshold $N$ (with some small unknown range around it),
there are \emph{two} estimated thresholds $N_1, N_2$,
such that $N_1 < N_2$,
with a some small unknown range around each of them, and
a \emph{large} difference between them. Therefore, there are
three main ranges of the numbers of photons $k$:
(a)~for $k \lesssim N_1$ photons, Bob's detectors work well
(and click if at least one photon arrives);
(b)~for $N_1 \lesssim k \lesssim N_2$ photons,
it would have become \emph{known} that some strange phenomena
happen---for example,
that Bob's detectors switch to the ``linear mode'';
and (c)~for $k \gtrsim N_2$ photons, Bob's detectors malfunction
(e.g., the detectors are burned).

Thus, surprisingly, even if the experimentalist had not known about the
two modes of operation (``Geiger mode'' and the ``linear mode'')
existing for each detector, he or she could still have discovered
the two different thresholds $N_1, N_2$ and then investigated
the detectors' behavior in the middle range
$N_1 \lesssim k \lesssim N_2$. This would have allowed him or her
to discover the ``linear mode'' and realize that there is also a need
to check \emph{multiple} pulses for finding the correct ``space of the
protocol'' and
for analyzing the security against ``Reversed-Space'' attacks.
Namely, the ``Reversed-Space'' approach makes it possible to discover
attacks even if the detectors are treated as \emph{a black box}
whose internal behavior is unknown.
By theoretically trying to prove security against any theoretical
``Reversed-Space'' attack, it would have been possible to find
the practical ``Bright Illumination'' attack; it would have even been
possible to study the operation of a ``\emph{black-box}'' detector and
discover, for example, that it has a ``linear mode'' of operation
(even if this mode of operation had not been
already known for realistic detectors).

\section{Conclusion}
We have seen a rare example (in quantum information processing)
where experiment preceded theory.
We can see now that this experimental attack could
have been theoretically predicted:
for a system to be secure, Bob must be sure that Eve cannot attack
by sending an unexpected number of photons, and he must know what
happens to his detectors for any number of photons.
Otherwise---Eve can attack;
and we could have known that this may be possible.
We have also defined the general notion of ``quantum side-channel
attacks'', distinguishing ``state-channel attacks''
(including ``Reversed-Space'' and ``Quantum Space'' attacks)
that interact with the transmitted (prepared or measured)
quantum states,
from ``quantum side-channel attacks'' that \emph{do not interact}
with the transmitted quantum states.

\bibliographystyle{splncs04}
\bibliography{Bright-Reversed}

\begin{thebibliography}{10}
\providecommand{\url}[1]{\texttt{#1}}
\providecommand{\urlprefix}{URL }
\providecommand{\doi}[1]{https://doi.org/#1}

\bibitem{bb84_experiment}
Bennett, C.H., Bessette, F., Brassard, G., Salvail, L., Smolin, J.:
  Experimental quantum cryptography. J. Cryptol.  \textbf{5}(1),  3--28 (1992),
  \url{https://link.springer.com/article/10.1007/BF00191318}

\bibitem{bb84}
Bennett, C.H., Brassard, G.: Quantum cryptography: Public key distribution and
  coin tossing. In: International Conference on Computers, Systems \& Signal
  Processing. pp. 175--179 (1984)

\bibitem{teleport}
Bennett, C.H., Brassard, G., Cr\'epeau, C., Jozsa, R., Peres, A., Wootters,
  W.K.: Teleporting an unknown quantum state via dual classical and
  {Einstein-Podolsky-Rosen} channels. Phys. Rev. Lett.  \textbf{70},
  1895--1899 (1993), \url{https://link.aps.org/doi/10.1103/PhysRevLett.70.1895}

\bibitem{fixed14}
Boyer, M., Gelles, R., Mor, T.: Attacks on fixed-apparatus
  quantum-key-distribution schemes. Phys. Rev. A  \textbf{90},  012329 (2014),
  \url{http://link.aps.org/doi/10.1103/PhysRevA.90.012329}

\bibitem{BLMS00}
Brassard, G., L\"utkenhaus, N., Mor, T., Sanders, B.C.: Limitations on
  practical quantum cryptography. Phys. Rev. Lett.  \textbf{85},  1330--1333
  (2000), \url{http://link.aps.org/doi/10.1103/PhysRevLett.85.1330}

\bibitem{BLMS00_eurocrypt}
Brassard, G., L{\"u}tkenhaus, N., Mor, T., Sanders, B.C.: Security aspects of
  practical quantum cryptography. In: Preneel, B. (ed.) Advances in Cryptology
  --- EUROCRYPT 2000, Belgium. Proceedings. pp. 289--299. Springer Berlin
  Heidelberg, Berlin, Heidelberg (2000),
  \url{http://link.springer.com/chapter/10.1007/3-540-45539-6_20}

\bibitem{deutsch85}
Deutsch, D.: Quantum theory, the {Church}-{Turing} principle and the universal
  quantum computer. P. Roy. Soc. Lond. A Mat.  \textbf{400}(1818),  97--117
  (1985), \url{http://rspa.royalsocietypublishing.org/content/400/1818/97}

\bibitem{reversed_thesis}
Gelles, R.: On the Security of Theoretical and Realistic Quantum Key
  Distribution Schemes. Master's thesis, Technion -- Israel Institute of
  Technology, Haifa, Israel (2008),
  \url{http://www.graduate.technion.ac.il/theses/Abstracts.asp?Id=24946}

\bibitem{qsa07}
Gelles, R., Mor, T.: Quantum-space attacks. arXiv preprint arXiv:0711.3019
  (2007), \url{https://arxiv.org/abs/0711.3019}

\bibitem{reversed16}
Gelles, R., Mor, T.: Reversed space attacks. arXiv preprint arXiv:1110.6573
  (2011), \url{http://arxiv.org/abs/1110.6573}

\bibitem{reversed12}
Gelles, R., Mor, T.: On the security of interferometric quantum key
  distribution. In: Dediu, A.H., Mart{\'i}n-Vide, C., Truthe, B. (eds.) Theory
  and Practice of Natural Computing 2012, Tarragona, Spain. Proceedings. pp.
  133--146. Springer Berlin Heidelberg, Berlin, Heidelberg (2012),
  \url{http://link.springer.com/chapter/10.1007/978-3-642-33860-1_12}

\bibitem{gllp04}
Gottesman, D., Lo, H.K., L{\"u}tkenhaus, N., Preskill, J.: Security of quantum
  key distribution with imperfect devices. Quantum Inf. Comput.  \textbf{4}(5),
   325--360 (2004),
  \url{http://www.rintonpress.com/journals/qiconline.html#v4n5}

\bibitem{decoy_hwang03}
Hwang, W.Y.: Quantum key distribution with high loss: Toward global secure
  communication. Phys. Rev. Lett.  \textbf{91},  057901 (2003),
  \url{https://link.aps.org/doi/10.1103/PhysRevLett.91.057901}

\bibitem{side_channel07}
K\"{o}pf, B., Basin, D.: An information-theoretic model for adaptive
  side-channel attacks. In: Proceedings of the 14th ACM Conference on Computer
  and Communications Security. pp. 286--296. CCS '07, Association for Computing
  Machinery, New York, NY, USA (2007),
  \url{https://dl.acm.org/doi/10.1145/1315245.1315282}

\bibitem{fixed_example02}
Kurtsiefer, C., Zarda, P., Halder, M., Weinfurter, H., Gorman, P.M., Tapster,
  P.R., Rarity, J.G.: A step towards global key distribution. Nature
  \textbf{419}(6906),  450--450 (2002),
  \url{https://www.nature.com/articles/419450a}

\bibitem{qkd_security_review}
Lo, H.K., Curty, M., Tamaki, K.: Secure quantum key distribution. Nat. Photon.
  \textbf{8}(8),  595--604 (2014),
  \url{http://www.nature.com/nphoton/journal/v8/n8/abs/nphoton.2014.149.html}

\bibitem{decoy_LMC05}
Lo, H.K., Ma, X., Chen, K.: Decoy state quantum key distribution. Phys. Rev.
  Lett.  \textbf{94},  230504 (2005),
  \url{https://link.aps.org/doi/10.1103/PhysRevLett.94.230504}

\bibitem{makarov10}
Lydersen, L., Wiechers, C., Wittmann, C., Elser, D., Skaar, J., Makarov, V.:
  Hacking commercial quantum cryptography systems by tailored bright
  illumination. Nat. Photon.  \textbf{4}(10),  686--689 (2010),
  \url{http://www.nature.com/nphoton/journal/v4/n10/abs/nphoton.2010.214.html}

\bibitem{faked_states06}
Makarov, V., Anisimov, A., Skaar, J.: Effects of detector efficiency mismatch
  on security of quantum cryptosystems. Phys. Rev. A  \textbf{74},  022313
  (2006), \url{http://link.aps.org/doi/10.1103/PhysRevA.74.022313}

\bibitem{faked_states05}
Makarov, V., Hjelme, D.R.: Faked states attack on quantum cryptosystems. J.
  Mod. Optic.  \textbf{52}(5),  691--705 (2005),
  \url{http://www.tandfonline.com/doi/abs/10.1080/09500340410001730986}

\bibitem{nv_teleport14}
Pfaff, W., Hensen, B.J., Bernien, H., van Dam, S.B., Blok, M.S., Taminiau,
  T.H., Tiggelman, M.J., Schouten, R.N., Markham, M., Twitchen, D.J., Hanson,
  R.: Unconditional quantum teleportation between distant solid-state quantum
  bits. Science  \textbf{345}(6196),  532--535 (2014),
  \url{http://science.sciencemag.org/content/345/6196/532}

\bibitem{quantum_computers_NISQ18}
Preskill, J.: Quantum computing in the {NISQ} era and beyond. Quantum
  \textbf{2}, ~79 (2018),
  \url{https://quantum-journal.org/papers/q-2018-08-06-79/}

\bibitem{SARG04}
Scarani, V., Ac\'{\i}n, A., Ribordy, G., Gisin, N.: Quantum cryptography
  protocols robust against photon number splitting attacks for weak laser pulse
  implementations. Phys. Rev. Lett.  \textbf{92},  057901 (2004),
  \url{https://link.aps.org/doi/10.1103/PhysRevLett.92.057901}

\bibitem{qkd_security_practical}
Scarani, V., Bechmann-Pasquinucci, H., Cerf, N.J., Du\ifmmode~\check{s}\else
  \v{s}\fi{}ek, M., L\"utkenhaus, N., Peev, M.: The security of practical
  quantum key distribution. Rev. Mod. Phys.  \textbf{81},  1301--1350 (2009),
  \url{http://link.aps.org/doi/10.1103/RevModPhys.81.1301}

\bibitem{qkd_black_paper}
Scarani, V., Kurtsiefer, C.: The black paper of quantum cryptography: Real
  implementation problems. Theor. Comput. Sci.  \textbf{560},  27--32 (2014),
  \url{https://www.sciencedirect.com/science/article/pii/S0304397514006938},
  theoretical Aspects of Quantum Cryptography --- celebrating 30 years of BB84

\bibitem{shamir90s}
Shamir, A.: personal communication

\bibitem{shor_algo94}
Shor, P.W.: Algorithms for quantum computation: discrete logarithms and
  factoring. In: Proceedings 35th Annual Symposium on Foundations of Computer
  Science. pp. 124--134 (1994),
  \url{http://ieeexplore.ieee.org/document/365700/}

\bibitem{shor_algo99}
Shor, P.W.: Polynomial-time algorithms for prime factorization and discrete
  logarithms on a quantum computer. SIAM Rev.  \textbf{41}(2),  303--332
  (1999), \url{http://epubs.siam.org/doi/10.1137/S0036144598347011}

\bibitem{large_pulse01}
Vakhitov, A., Makarov, V., Hjelme, D.R.: Large pulse attack as a method of
  conventional optical eavesdropping in quantum cryptography. J. Mod. Opt.
  \textbf{48}(13),  2023--2038 (2001),
  \url{https://www.tandfonline.com/doi/abs/10.1080/09500340108240904}

\bibitem{decoy_wang05}
Wang, X.B.: Beating the photon-number-splitting attack in practical quantum
  cryptography. Phys. Rev. Lett.  \textbf{94},  230503 (2005),
  \url{https://link.aps.org/doi/10.1103/PhysRevLett.94.230503}

\end{thebibliography}

\end{document}